\documentclass{moriond}

\def\Journal#1#2#3#4{{#1} {\bf #2}, #3 (#4)}

\def\PLB{{\em Phys. Lett.}  B}
\def\PRL{\em Phys. Rev. Lett.}

\def\EPJC{{\em Eur. Phys. J.} C}

\def\be{\begin{equation}}
\def\ee{\end{equation}}
\def\bea{\begin{eqnarray}}
\def\eea{\end{eqnarray}}

\newcommand{\Photo}{}

\begin{document}
\vspace*{4cm}
\title{Rare and Exotic Higgs decays at ATLAS and CMS}

\author{ Pallabi Das on behalf of the ATLAS and CMS Collaborations \footnote{Copyright 2024 CERN for the benefit of the ATLAS and CMS Collaborations. Reproduction of this article or parts of it is allowed as specified in the CC-BY-4.0 license}}

\address{Department of Physics, Princeton University\\
Princeton, New Jersey 08544, USA}

\maketitle\abstracts{
After the Higgs boson discovery in 2012, the experiments at the LHC are continuing to study this particle and look for physics beyond the standard model. Some of the Higgs boson properties, such as the mass, has been measured with sub-percent level accuracy. Yet the present integrated luminosity is still a limiting factor for measuring the Higgs boson self-coupling or the first generation Yukawa couplings. The current constraints on the Higgs boson couplings would still allow for a sizeable branching fraction into undetected final states, which motivates the direct searches for rare and exotic decay modes. This presentation discusses several new results from these searches utilizing advanced online selection methods or analysis techniques with the entire Run 2 data.
}

\section{Introduction}
The ATLAS~\cite{atlas_paper} and CMS~\cite{cms_paper} experiments discovered the Higgs boson in Run 1 of the LHC~\cite{Aad:2012tfa,Chatrchyan:2012xdj,Chatrchyan:2013lba}. While testing the validity of the Standard Model (SM), the experiments also search for beyond SM (BSM) physics. However, with no evidence of new particles, understanding the nature of the Higgs boson is now of paramount importance. Several theories suggest that new physics particles couple preferentially to the Higgs boson. Even as the experiments are limited by integrated luminosity to probe several SM Higgs boson couplings, the available data includes potentially unexplored phase space. Direct searches of rare and exotic Higgs boson decays are designed to fully exploit the LHC collision data to examine the possibility of BSM interactions.

\section{Rare Higgs boson decays to a meson and a photon}
The existence of new physics may alter some very rare decay widths of the Higgs boson, e.g., processes that are suppressed in the SM, producing flavour changing neutral currents (FCNCs) or lepton flavour violation (LFV). Exclusive Higgs boson decays to a photon and a quarkonium state are also sensitive to Yukawa couplings of the charm and lighter quarks, the latter being out of reach with the present LHC dataset. Three new results are discussed in this presentation which reconstruct meson-photon final states resulting from Higgs boson decays. For all three searches, upper limits on the respective Higgs boson rare decay branching fraction are obtained by a fit of the meson-photon system invariant mass distribution.

\begin{figure}[!htb]
	\begin{center}
		\begin{tabular}{cc}
			{\includegraphics[width=2in]{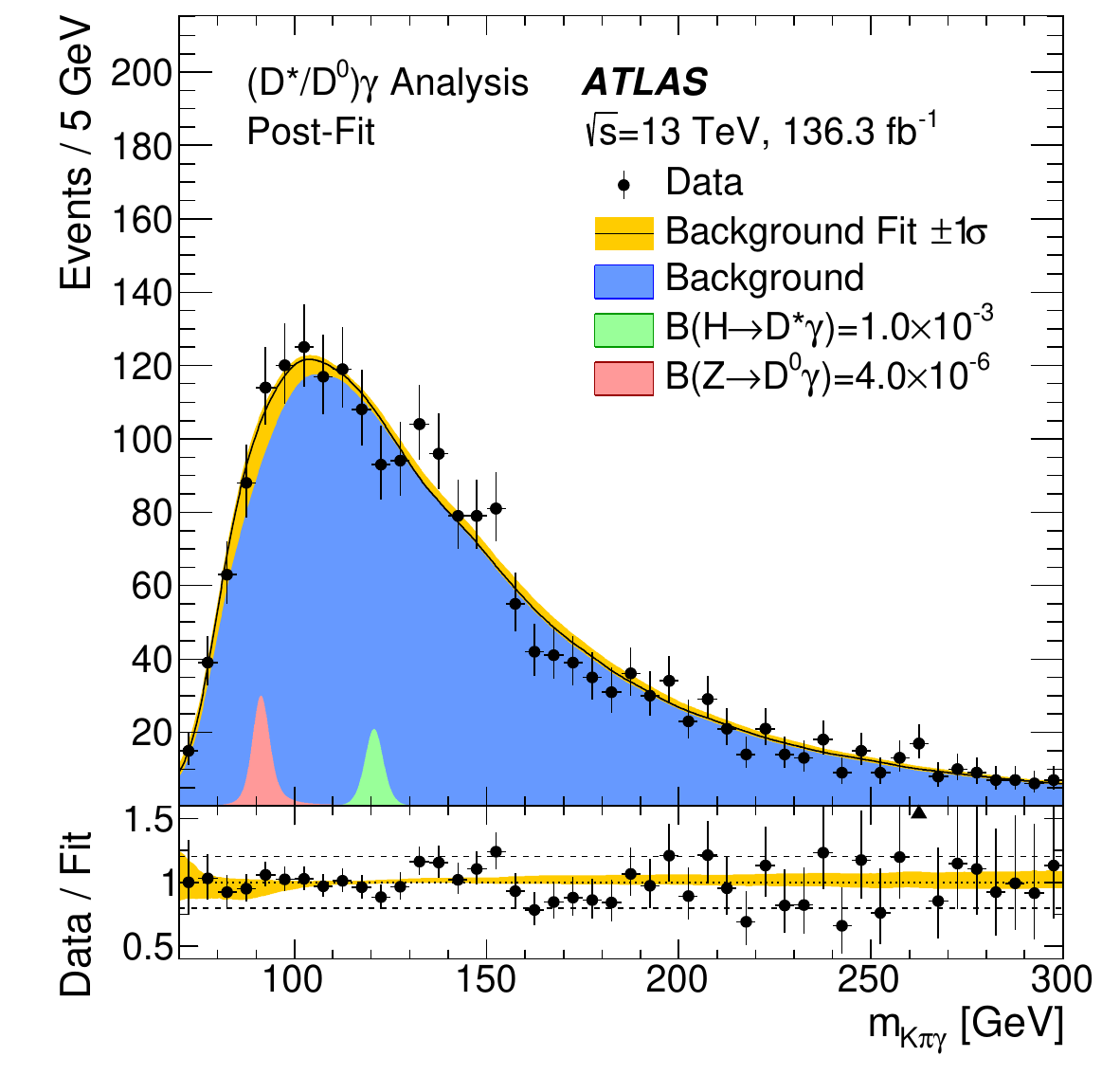}}
			{\includegraphics[width=2in]{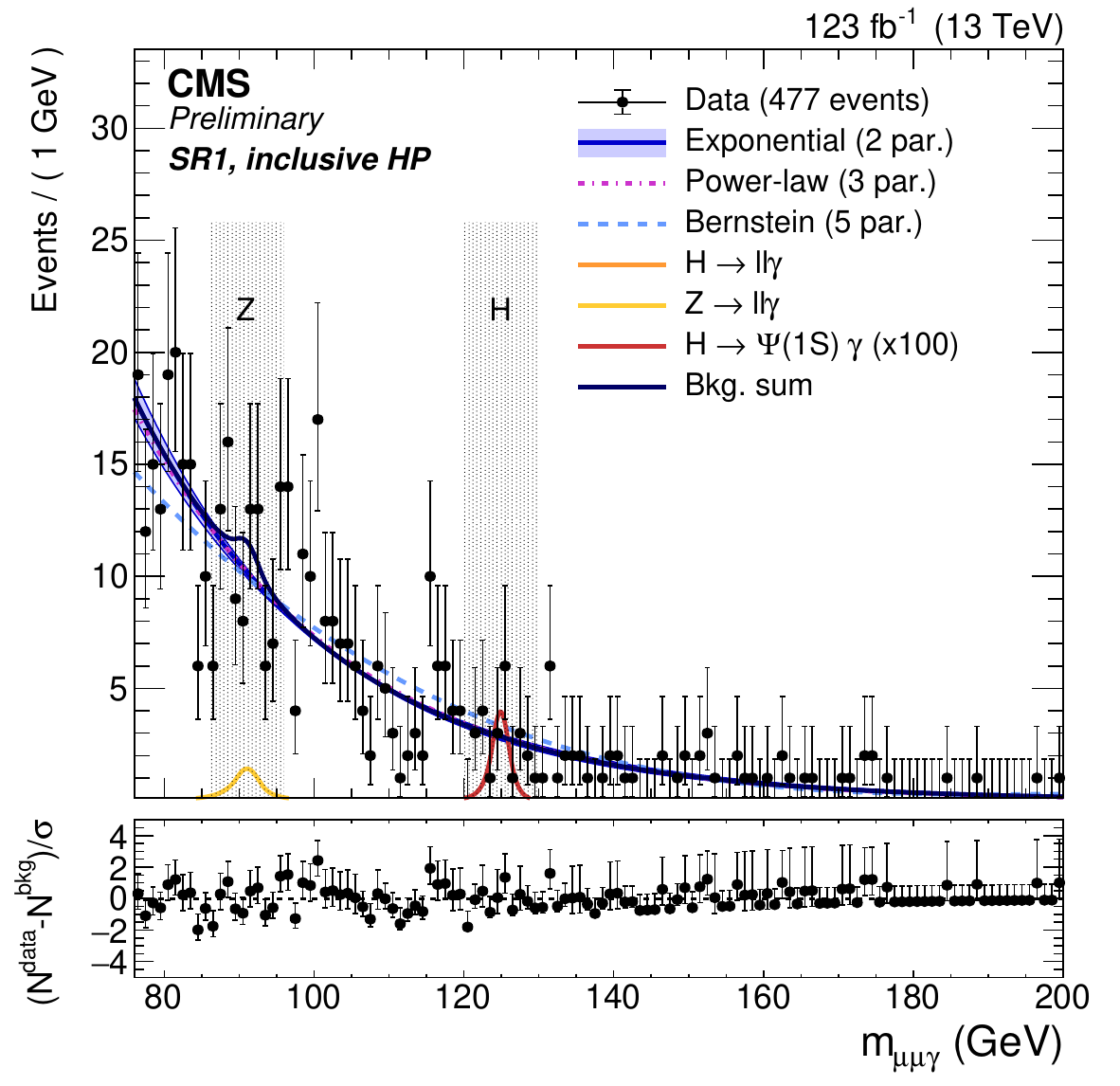}}
			{\includegraphics[width=1.9in]{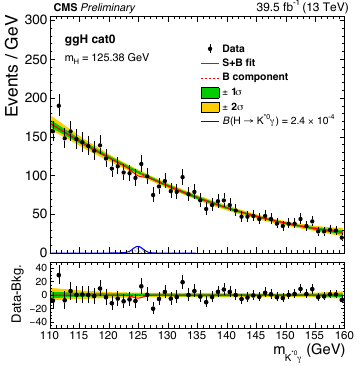}}
		\end{tabular}
	\end{center}
	\vspace*{-0.2in}
	\caption{Invariant mass spectra in the signal regions searching for (left) $\mathrm{H}\to D^*\gamma$~{\protect\cite{HDBS-2018-52}}, (middle) $\mathrm{H}\to \mathrm{J}/\psi\gamma$ or  $\psi'\gamma$~{\protect\cite{CMS-PAS-SMP-22-012}}, and (right) $\mathrm{H}\to K^{*0}\gamma$~{\protect\cite{CMS-PAS-HIG-23-005}}. }
	\label{fig1}
\end{figure}

The ATLAS experiment performed a search for $\mathrm{H}\to D^*\gamma$ which has an extremely low expected branching fraction of $\sim10^{-27}$ due to FCNC~\cite{HDBS-2018-52}. Dedicated online selection (trigger) is used to select events containing $D^0\to K^- \pi^+$ signature and offline requirements are placed on reconstructing the displaced decay vertex of the $D^0$ meson. The first ever 95\% CL upper limit on $\mathcal{B}(\mathrm{H}\to D^*\gamma)$ is obtained with observed values below $10^{-3}$. The analysis also constrains $\mathcal{B}(\mathrm{Z}\to K^{s0}\gamma)$ and $\mathcal{B}(\mathrm{Z}\to D^0\gamma)$ values. The invariant mass distribution for the data and background model in the signal region of this analysis are shown in Fig.~\ref{fig1} (left).

The CMS search for $\mathrm{H}\to \mathrm{J}/\psi\gamma$ or  $\psi'\gamma$ is sensitive to charm Yukawa coupling~\cite{CMS-PAS-SMP-22-012}. The 95\% CL allowed range of the coupling modifier ratio $\kappa_c/\kappa_{\gamma}$ is within (-157, 199). The upper limits on Z boson branching fractions are also obtained. The photon-meson invariant mass spectrum is shown in Fig.~\ref{fig1} (middle). Observed $\mathcal{B}(\mathrm{H}\to J/\psi\gamma)$  and $\mathcal{B}(\mathrm{H}\to \psi'\gamma)$ values are constrained below $9.9\times10^{-4}$ and $2.6\times10^{-4}$, respectively, at 95\% CL.

Very recently, the CMS experiment has probed the rare final states of $\mathrm{H}\to \rho^0\gamma$ or $\phi\gamma$ or $K^{*0}\gamma$~\cite{CMS-PAS-HIG-23-005}. Specifically, $\mathcal{B}(\mathrm{H}\to K^{*0}\gamma)$ is expected to be $\sim10^{-19}$ due to flavor violating couplings of the s and d quarks. While full Run 2 luminosity is analyzed, partial data was selected with a dedicated trigger requiring a single photon and meson event signature. This analysis provides the most stringent upper limits to date on these final states with observed values $\mathcal{B}(\mathrm{H}\to \rho^0\gamma)<3.74\times10^{-4}$,  $\mathcal{B}(\mathrm{H}\to \phi\gamma)<2.97\times10^{-4}$ and $\mathcal{B}(\mathrm{H}\to \rho^0\gamma)<1.71\times10^{-4}$ at 95\% CL. The Kaon-photon invariant mass spectrum in one of the analysis categories is shown in Fig.~\ref{fig1} (right).

\section{Higgs boson decays to dark photons}
\begin{figure}[!htb]
	\begin{center}
		\begin{tabular}{cc}
			{\includegraphics[width=3in]{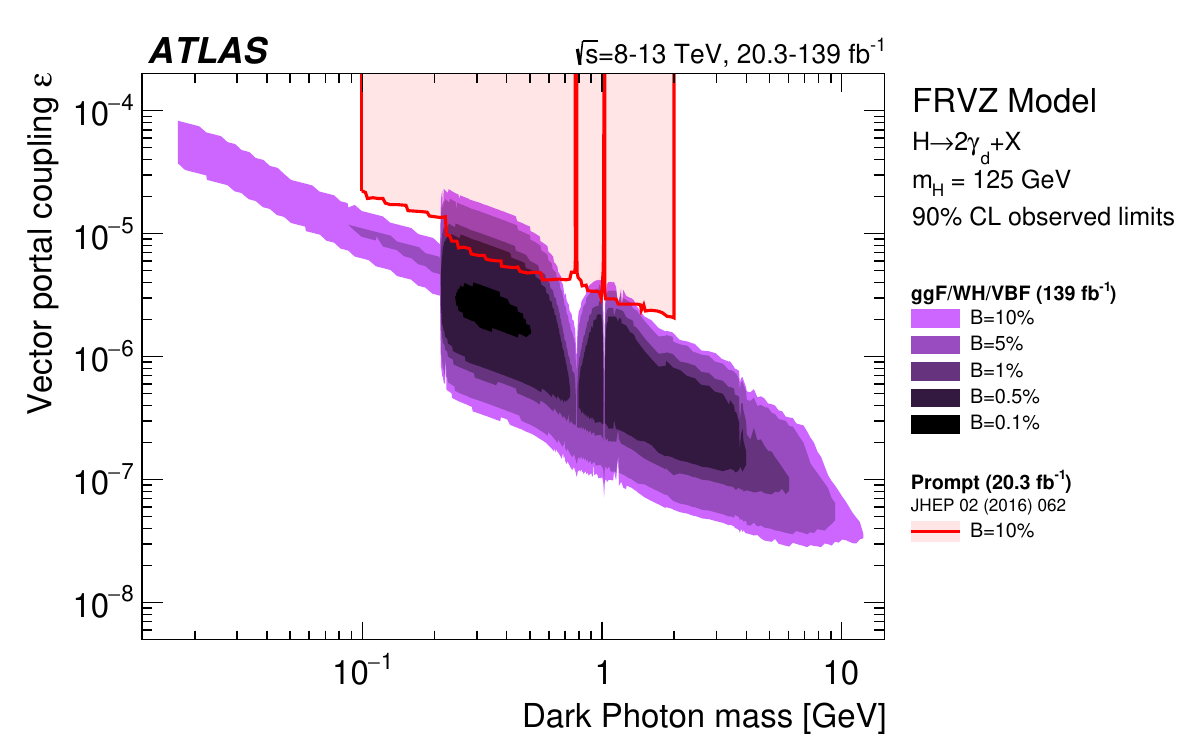}}
		\end{tabular}
	\end{center}
	\vspace*{-0.2in}
	\caption{90\% CL observed limits  on $\mathcal{B}(\mathrm{H}\to\gamma_d\gamma_d)$ considering both prompt and long-lived $\gamma_d$~{\protect\cite{EXOT-2022-15}}. }
	\label{fig2}
\end{figure}

An example of dark matter particles coupling preferentially to the Higgs boson, described in the FRVZ model~\cite{frvz1,frvz2}, is examined by the ATLAS experiment~\cite{EXOT-2022-15}. A search for Higgs boson decays to a pair of dark photons ($\gamma_d$) is performed considering $\gamma_d$ masses between 0.1--30 GeV. The lifetime of $\gamma_d$  depends on the kinematic mixing parameter, $\epsilon$, with SM photon. A specially designed trigger targets long-lived $\gamma_d$ signature and vector-boson-fusion Higgs production mode. The displaced collimated shower of SM particles is reconstructed in the calorimeter and the muon detectors. 
Figure~\ref{fig2} shows the 90\% CL observed limits on $\mathcal{B}(\mathrm{H}\to\gamma_d\gamma_d)$ as a function of the $\gamma_d$ proper decay length and  $\epsilon$. Values above 10\% are excluded for a mass of 10 GeV and mean proper decay length between 173--1296 mm corresponding to small values of $\epsilon$ ($<10^{-5}$).

\section{Higgs boson decays to axion-like particles}
Axion-like particles or ALPs are gauge-singlet pseudoscalars predicted in extended SM theories to address the strong CP problem. ALP decays to diphoton can be identified at the LHC experiments and provide complimentary measurements to cosmological searches. The ATLAS experiment has recently studied the on-shell Higgs boson decay modes $\mathrm{H}\to \mathrm{aa}$ and  $\mathrm{H}\to \mathrm{Za}$, with $\mathrm{a}\to\gamma\gamma$, using the $4\gamma$ and diphoton invariant mass distributions, respectively. The photon signature may be merged for low mass of the ALPs, reconstructed as a single object, or fully resolved.

\begin{figure}[!htb]
	\begin{center}
		\begin{tabular}{cc}
			{\includegraphics[width=2.8in]{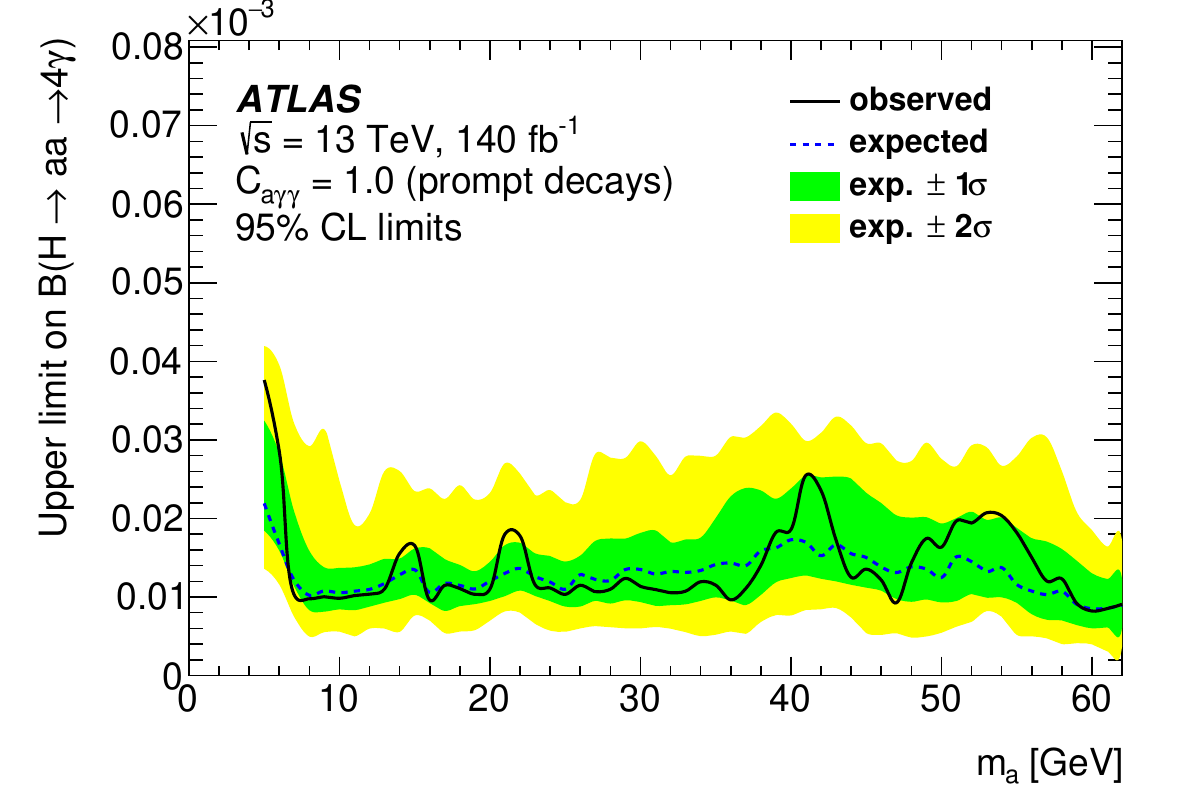}}
			{\includegraphics[width=1.96in]{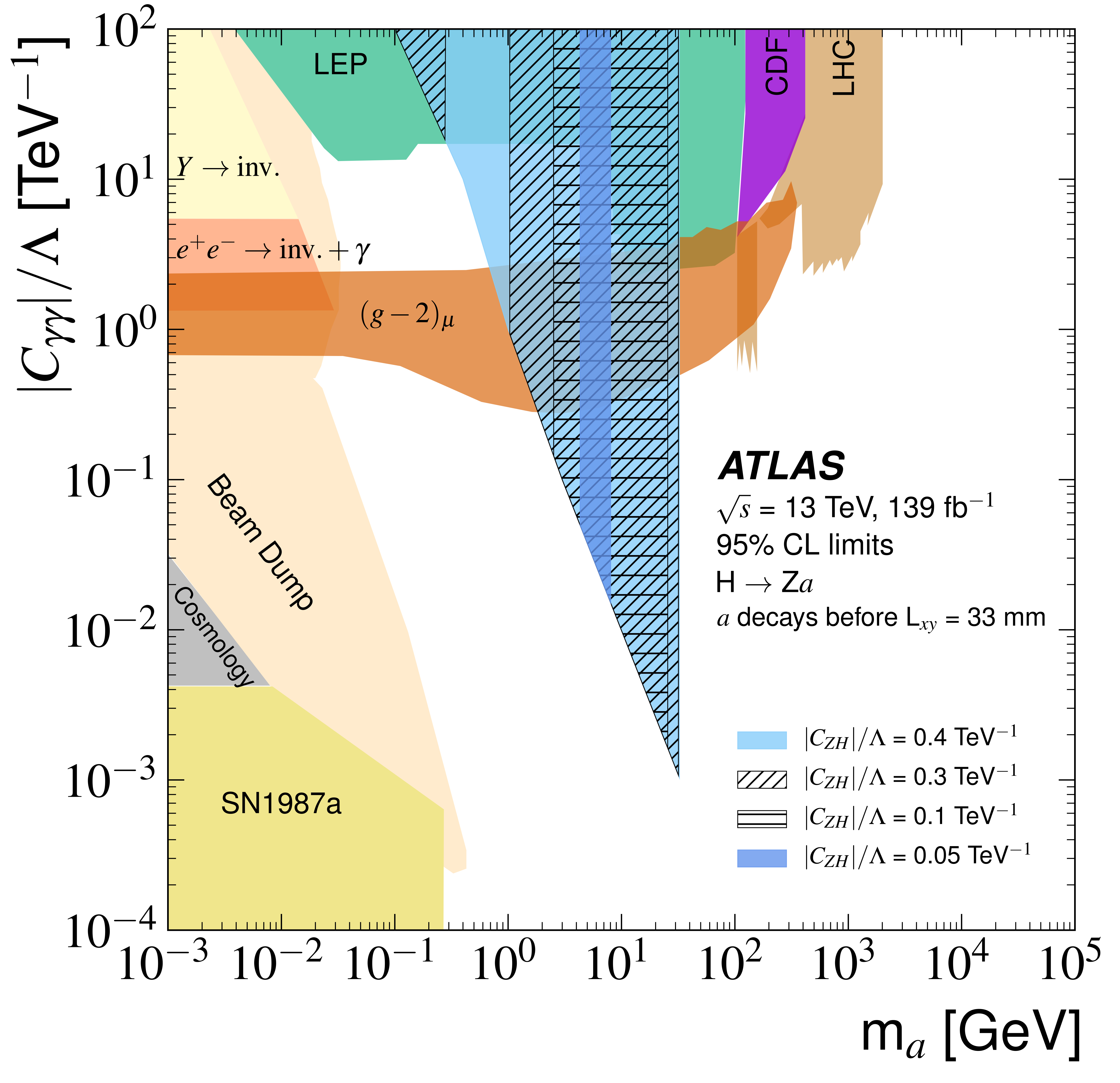}}
		\end{tabular}
	\end{center}
	\vspace*{-0.2in}
	\caption{95\% CL upper limits  on (left) $\mathcal{B}(\mathrm{H}\to\mathrm{aa}\to 4\gamma)$~{\protect\cite{HDBS-2019-19}} and (right) effective ALP coupling to photons~{\protect\cite{htoza}} as a function of the ALP mass. }
	\label{fig3}
\end{figure}

In the search for $\mathrm{H}\to \mathrm{aa}\to 4\gamma$, a mass range of $0.1<m_{\mathrm{a}}<60$ GeV is considered for both prompt and long-lived signatures~\cite{HDBS-2019-19}. This analysis provides the most stringent upper limits till date on long-lived ALPs with observed branching fractions $<$ 2--6$\times10^{-5}$ for $m_{\mathrm{a}} > 10$ GeV and $<10^{-5}$--3$\times10^{-2}$ for $m_{\mathrm{a}} < 10$ GeV at 95\% CL. Upper limits on prompt signatures are illustrated in Fig.~\ref{fig3} (left). Only promptly decaying ALPs are considered in the search for $\mathrm{H}\to \mathrm{Za} \to \ell\ell\gamma\gamma$~\cite{htoza}. $\mathcal{B}(\mathrm{H}\to \mathrm{Za})$ values above 0.08--2$\times10^{-2}$ are excluded at 95\% CL within a mass range $0.1<m_{\mathrm{a}}<33$ GeV. The upper limits on the effective couplings of the ALP to photons, $|C_{\gamma\gamma}|/\Lambda$, are shown in Fig.~\ref{fig3} (right).
\section{Higgs boson decays to a pair of pseudoscalars}
Several other beyond SM theories predict Higgs boson exotic decays to a pair of pseudoscalars. They occur naturally in two-Higgs-doublet models extended with a singlet (2HDM+S), which is a special case of next-to-minimal supersymmetric model (NMSSM). Other possibilities include supersymmetric (SUSY) models with hidden sectors (dark-SUSY), vector-portal models etc. Many possible final states from $\mathrm{H}\to\mathrm{aa}$ decays are experimentally probed by the CMS and ATLAS experiments. In this presentation two recent CMS searches with the full Run 2 data set are reported.

\begin{figure}[!htb]
	\begin{center}
		\begin{tabular}{cc}
			{\includegraphics[width=2in]{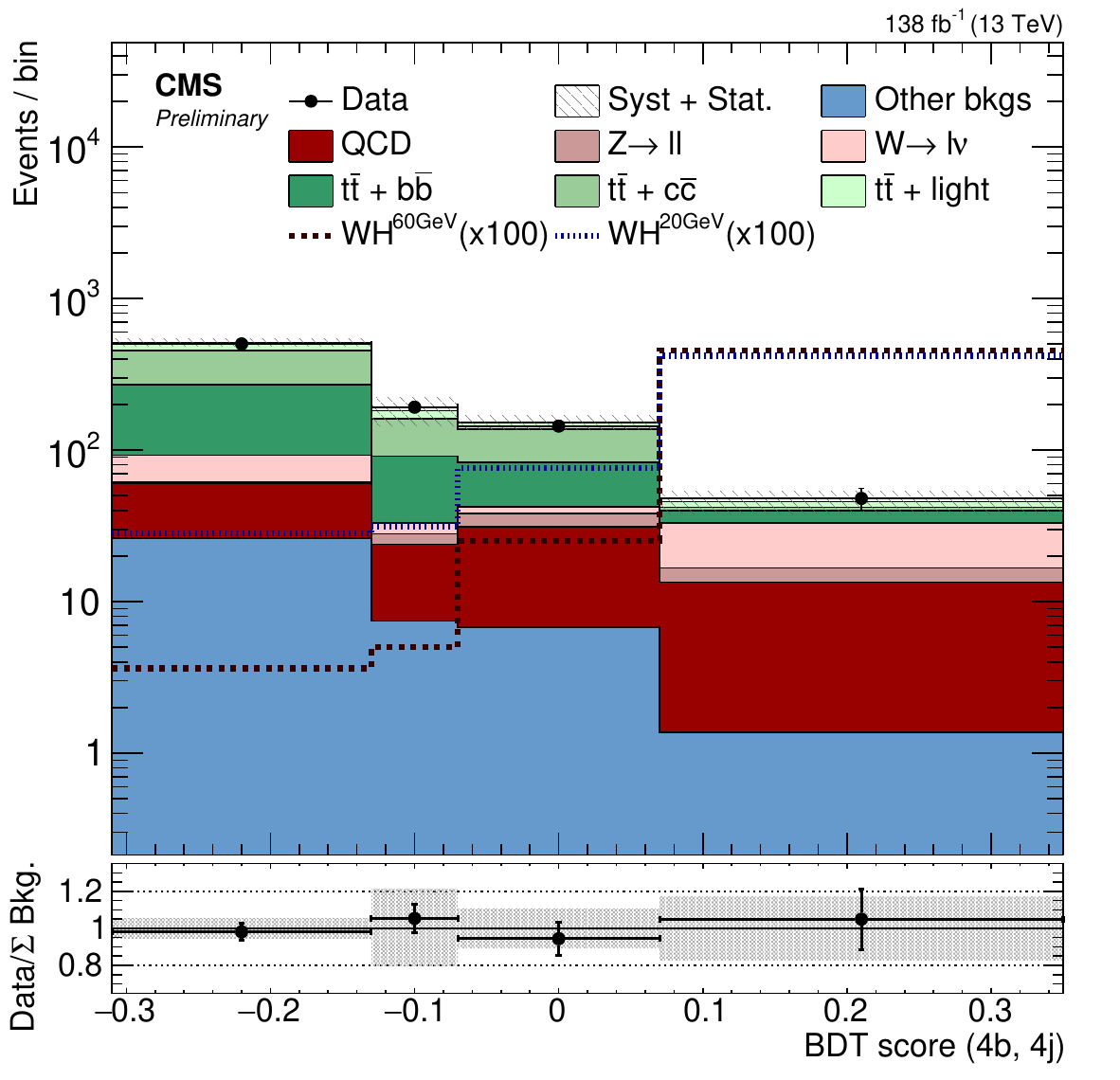}}
			{\includegraphics[width=1.9in]{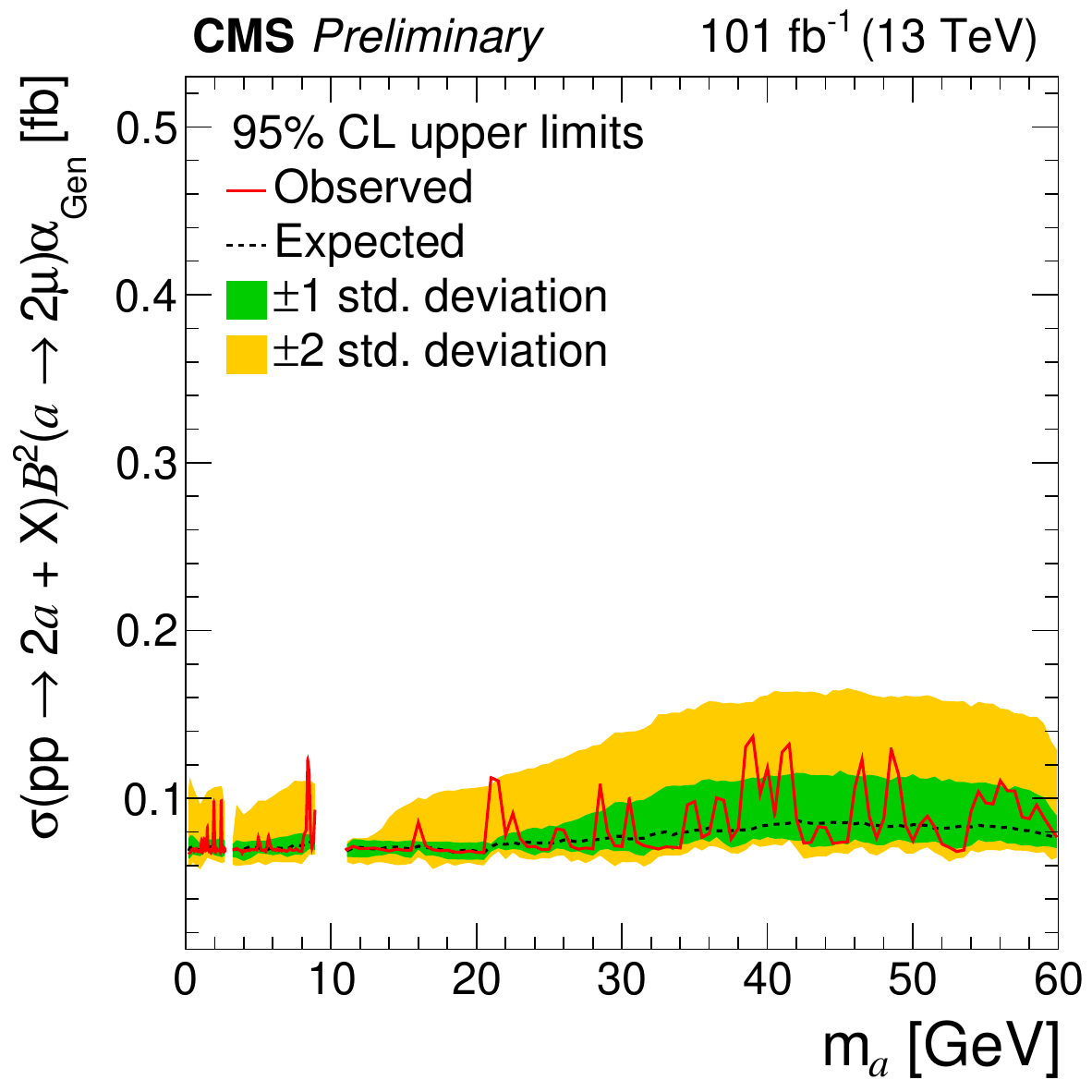}}
		\end{tabular}
	\end{center}
	\vspace*{-0.2in}
	\caption{(Left) BDT score distribution in one of the signal regions of $\mathcal{B}(\mathrm{H}\to\mathrm{aa}\to 4\mathrm{b})$ analysis~{\protect\cite{CMS-PAS-HIG-18-026}} and (right) 95\% CL upper limits on $\sigma(\mathrm{H}\to2\mathrm{a+X})\times\mathcal{B}^2(\mathrm{a}\to2\mu)$ as a function of the pseudoscalar mass~{\protect\cite{CMS-PAS-HIG-21-004}}. }
	\label{fig4}
\end{figure}

In the first search with both pseudosclars decaying to a pair of b-quarks, the vector-boson-associated production of the Higgs boson is considered which can be selected using leptonic trigger~\cite{CMS-PAS-HIG-18-026}. The fully hadronic final state is difficult to reconstruct, and the analysis searches for a signal in different categories based on the number of jets and further b-tagging requirements. A boosted decision tree (BDT) is used to discriminate the signal from background and the score is used as the discriminating variable for signal extraction. An example of the BDT score distribution is shown in Fig.~\ref{fig4} (left). Assuming $\mathcal{B}(\mathrm{H}\to\mathrm{aa}\to4\mathrm{b})$ to be 100\%, the analysis excludes $m_{\mathrm{a}}$ values between 21--60 GeV at 95\% CL.
	
The second search probes the $\mathrm{H}\to\mathrm{aa}\to4\mu$ final state considering $0.21 <m_{\mathrm{a}} < 60$ GeV and proper decay length between 0--100 mm~\cite{CMS-PAS-HIG-21-004}. The results are interpreted in terms of vector-portal models, ALPs, NMSSM and dark-SUSY scenarios. A long-lived muon trigger is used for selecting events, but this trigger was only active in part of the Run 2 data set. Since both muon systems can be well reconstructed, signal selection is based on the agreement between the dimuon mass and the mass of the pseudoscalar. The 95\% CL observed upper limit on $\sigma(\mathrm{H}\to2\mathrm{a+X})\times\mathcal{B}^2(\mathrm{a}\to2\mu)$ is within the range 0.049--0.247 fb. The model independent upper limits are shown in Fig.~\ref{fig4} (right).

\section {Summary}
The 125 GeV Higgs boson serves as a probe for dark matter candidates and other exotic particles from extended Higgs sectors. On the other hand, exclusive Higgs boson decays to a final state with quarkonia can probe anomalous Yukawa couplings to charm and lighter quarks. While the searches for rare and exotic Higgs boson decays are dominated by statistical uncertainties, they are sensitive to the Higgs boson interaction with new physics. No excess over SM is yet observed, but the LHC experiments are working on improving search strategies and exploring all available phase space. Going forward, Run 3 of LHC will increase BSM signal sensitivity by adding new phase space in terms of data volume and specially designed triggers.

\end{document}